\documentclass[aps,prl,reprint,superscriptaddress,floatfix,nofootinbib,showkeys,showpacs,preprintnumbers]{revtex4-1}

\pdfoutput=1

\usepackage{tabularx}
\usepackage{xspace}
\usepackage{listings}
\usepackage{lineno}

\usepackage[dvipsnames,svgnames]{xcolor} 
\usepackage{graphicx}
\usepackage{comment}

\usepackage{natbib}
\usepackage{hyperref}
\usepackage{amsmath}
\usepackage{enumitem}
\setlist{wide, labelwidth=!, labelindent=0pt}
\usepackage{multirow}
\hypersetup{    
  colorlinks      = true,
  linkcolor       = {blue},
  linkbordercolor = {white},
  citecolor       = {blue},
  citebordercolor = {white},
  urlcolor        = {blue},
  urlbordercolor  = {white},
}

\usepackage{lipsum}
\usepackage{aas_macros}
\usepackage{orcidlink}

\begin{document}

\title{Constraints on the epoch of dark matter formation from Milky Way satellites}

\author{Subinoy~Das\orcidlink{0000-0002-7771-180X}}
\email[]{subinoy@iiap.res.in}
\affiliation{Indian Institute of Astrophysics, Sarjapura Road, 2nd Block Koramangala, Bengaluru, Karnataka, 560034, India}

\author{Ethan~O.~Nadler\orcidlink{0000-0002-1182-3825}}
\email[]{enadler@stanford.edu}
\affiliation{Kavli Institute for Particle Astrophysics and Cosmology and Department of Physics, Stanford University, Stanford, California 94305, USA}

\date{\today}

\begin{abstract}
A small fraction of thermalized dark radiation that transitions into cold dark matter (CDM) between big bang nucleosynthesis and matter-radiation equality can account for the entire dark matter relic density. Because of its transition from dark radiation, ``late-forming dark matter'' (LFDM) suppresses the growth of linear matter perturbations and imprints the oscillatory signatures of dark radiation perturbations on small scales. The cutoff scale in the linear matter power spectrum is set by the redshift $z_T$ of the phase transition; tracers of small-scale structure can therefore be used to infer the LFDM formation epoch. Here, we use a forward model of the Milky Way (MW) satellite galaxy population to address the question: How late can dark matter form? For dark radiation with strong self-interactions, which arises in theories of neutrinolike LFDM, we report~$z_{T}>5.5\times 10^6$ at~$95\%$ confidence based on the abundance of known MW satellite galaxies. This limit rigorously accounts for observational incompleteness corrections, marginalizes over uncertainties in the connection between dwarf galaxies and dark matter halos, and improves upon galaxy clustering and Lyman-$\alpha$ forest constraints by nearly an order of magnitude. We show that this limit can also be interpreted as a lower bound on $z_T$ for LFDM that free-streams prior to its phase transition, although dedicated simulations will be needed to analyze this case in detail. Thus, dark matter created by a transition from dark radiation must form no later than one week after the big bang.
\end{abstract}

\maketitle


\section{I.\ Introduction}
\label{sec:intro}

Despite intensive experimental searches in recent decades, the nature of dark matter (DM) remains a mystery. Combined with a cosmological constant ($\Lambda$), the simple hypothesis of a cold, collisionless dark matter (CDM) particle that interacts extremely weakly with Standard Model (SM) particles is consistent with all cosmological observations to date, on scales ranging from individual galaxies \cite{Rubin:1970zza}, to galaxy clusters \cite{Zwicky:1933gu}, to the cosmological horizon as probed by large-scale structure~\cite{Ata:2017dya} and cosmic microwave background (CMB) measurements~\cite{2016A&A...596A.107P,Bertone:2004pz}. However, particle physics experiments have not detected canonical weakly interacting-massive-particle (WIMP) CDM, and several astrophysical anomalies have been claimed to provide evidence for physics beyond the collisionless CDM paradigm \cite{Bullock170704256}.

In this work, we explore and strongly constrain one such alternative scenario, known as ``late-forming dark matter'' (LFDM), where DM appears much later in cosmic history than WIMPs and other popular DM candidates \cite{Das:2006ht,Das200511889}. Instead of focusing on a specific particle physics construction of LFDM, we consider a general class of models in which DM is produced from an excess (dark) radiation component that undergoes a phase transition due to nontrivial interactions in the dark sector. Measurements from the \emph{Planck} mission rule out the existence of a fully thermalized extra radiation component during the epoch of the CMB \cite{2016A&A...596A.107P}. However, as we will demonstrate, LFDM can account for the entire DM content of the Universe while remaining compatible with \emph{Planck} limits on the number of excess light degrees of freedom if even a tiny fraction of dark radiation transitions into CDM between the epoch of big bang nucleosynthesis (BBN) and the CMB. 

LFDM is intriguing because it can be realized as a light, neutrinolike particle \cite{Das:2006ht,Das200511889}, reviving the possibility of $\sim \mathrm{eV}$-mass neutrinolike DM, which is incompatible with structure formation constraints if produced thermally \cite{Bell:2005dr,Acero:2008rh,Aghanim:2018eyx,Nadler200800022}. Intriguingly, there are tentative hints of a fourth sterile neutrino generation from short-baseline neutrino oscillation experiments \cite{MiniBooNE2010,MiniBooNE2013,MiniBooNE1805}. However, this signal does not appear ubiquitously (e.g., \cite{Bay200200301}) and its interpretation as a sterile neutrino is difficult to reconcile with cosmological observables (e.g., \cite{Hagstotz200302289}). Moreover, within the ``3+1'' neutrino oscillation framework, these results are difficult to reconcile with the absence of anomalies in $\nu_{\mu}$ disappearance as probed by recent atmospheric \cite{Aartsen:2017bap,TheIceCube:2016oqi} and short-baseline \cite{Adamson:2017zcg,Aartsen:2017bap,An:2012eh} experiments. Thus, if the existence of a fourth sterile neutrino generation is confirmed by future analyses, it is likely that new physics beyond sterile-plus-active oscillation models is necessary to resolve the tension between neutrino appearance and disappearance data. Whether LFDM models can be connected to these anomalies is a compelling question for sterile neutrino model building, and is not the aim of this paper. Instead, we focus on cosmological signatures of the LFDM phase transition.

The LFDM phase transition affects linear matter perturbations and imprints its effects on various tracers of the DM density field throughout cosmic history. In particular, the linear matter power spectrum $P(k)$ is suppressed on scales smaller than the size of the cosmological horizon at the LFDM transition redshift, $z_T$, because the corresponding modes entered the horizon while LFDM behaved like radiation. Thus, later phase transitions suppress power on larger scales. This phenomenology pertains to \emph{any} cosmic fluid that transitions into CDM from a (dark) radiation component. Moreover, because the absence of cold, heavy DM particles always dilutes gravitational potentials, it also pertains to any scenario in which DM is absent until late times. 

In this work, we leverage this power suppression signal to address the question: ``What is the latest epoch after which dark matter must behave exactly like CDM?'' We show that the answer depends on whether the LFDM fluid has strong self-interactions prior to its transition into CDM (we refer to this case as \emph{self-interacting}, or SI), or whether it free-streams prior to the phase transition (we refer to this case as \emph{free-streaming}, or FS). In the SI LFDM case, the linear matter power spectrum contains the oscillatory signatures of dark radiation perturbations, the amplitude of which depends on the strength of the LFDM self-interactions prior to the phase transition \cite{Das:2006ht,Das:2018ons}. These self-interactions are expected in neutrinolike LFDM models, including in theories of neutrino dark energy \cite{Das:2006ht} and in a model of sterile fermion DM that has been proposed to have some observable effects on CMB ~\cite{Das200511889}. Meanwhile, the limit in which LFDM transitions to CDM from a free-streaming dark radiation component without self-interactions yields a sharper cutoff in the matter power spectrum.

Analyses of the Lyman-$\alpha$ forest, galaxy clustering, and the high-redshift galaxy luminosity function have set a lower limit on the SI LFDM transition redshift of~$z_{T,\mathrm{SI}}\gtrsim 9\times 10^{5}$ based on the lack of observed power spectrum suppression relative to CDM on quasilinear scales corresponding to wave numbers~$k\sim 1 h\ \mathrm{Mpc}^{-1}$~\cite{Sarkar14107129,Corasaniti161105892}. Following the reasoning above, tracers of matter fluctuations on even smaller scales contain information about earlier LFDM transition redshifts. Indeed, LFDM initially gained popularity because of its ability to address several ``small-scale crises'' historically attributed to CDM, including the ``missing satellites''~\cite{Klypin9901240,Moore9907411} and ``too big to fail'' \cite{Boylan-Kolchin11030007,Garrison-Kimmel14045313} problems for Milky Way (MW) satellite galaxies, which occupy DM halos that arise from fluctuations on nonlinear scales of~$k\gtrsim 10 h\ \mathrm{Mpc}^{-1}$.

State-of-the-art empirical models \cite{Jethwa161207834,Newton170804247,Kim:2018,Nadler180905542,Nadler191203303} and hydrodynamic simulations \cite{Sawala151101098,Wetzel160205957,Garrison-Kimmel180604143} combined with rigorous estimates for the incompleteness of current MW satellite searches provide strong evidence that the observed MW satellite population is consistent with CDM predictions. Recently, \cite{Nadler200800022} used the MW satellite model in~\cite{Nadler180905542,Nadler191203303}---which accurately describes the observed MW satellite population over nearly three-fourths of the sky, including satellites associated with the Large Magellanic Cloud---to derive constraints on a variety of non-CDM models that suppress the linear matter power spectrum on small scales. In particular, \cite{Nadler200800022} reported that the observed MW satellite population is consistent with CDM predictions down to a halo mass scale of~$\sim 3\times 10^8\ \mathrm{M}_{\mathrm{\odot}}$, corresponding to characteristic wave numbers~$k\sim 40 h\ \mathrm{Mpc}^{-1}$, and ruled out thermal relic warm dark matter (WDM) lighter than $6.5\ \mathrm{keV}$ at $95\%$ confidence. Importantly, this constraint is marginalized over uncertainties in the connection between faint galaxies and low-mass halos and the properties of the MW system. Independent studies of other small-scale structure probes, including the Lyman-$\alpha$ forest, strong gravitational lenses, and stellar streams, have derived consistent WDM constraints \cite{Viel:2013,Irsic:2017,Hsueh190504182,Gilman190806983,Banik191102662}.

Here, we extend the analysis of \cite{Nadler200800022} to place limits on the LFDM formation epoch. We show that SI LFDM imprints a cutoff in the linear matter power spectrum that is very similar to thermal relic WDM, and we exploit this correspondence to constrain the model. Based on the abundance of MW satellite galaxies, our analysis yields a lower bound of $z_{T,\mathrm{SI}}>5.5\times 10^6$ on the SI LFDM transition redshift at $95\%$ confidence, which improves upon previous results~\cite{Sarkar14107129,Corasaniti161105892} by a factor of~$\sim 6$. This implies that SI LFDM must form no later than one week after the big bang. In addition, we show that our constraint on $z_{T,\mathrm{SI}}$ can be interpreted as a lower limit on the FS LFDM transition redshift, and we estimate the improvement that future simulation-based analyses can provide for this model.

Throughout, we assume that LFDM constitutes the entire DM relic density, and we hold cosmological parameters fixed at the $\Lambda$CDM best-fit values from \cite{Ade:2015xua}.


\section{II.\ Late-forming Dark Matter Models}
\label{sec:model_fs}

We begin with a brief overview of LFDM physics. We consider LFDM models in which an excess radiation component $\Delta N_{\rm eff}$ undergoes a phase transition to a CDM state at redshift $z_T$. In this scenario, the initial number of relativistic degrees of freedom $N_{\rm eff}$ is generically larger than in a standard $\Lambda$CDM cosmology. However, we will see that even a tiny fractional increase in $N_{\rm eff}$ suffices to produce the observed CDM relic density, provided that the LFDM phase transition occurs a few $e$-foldings before matter-radiation equality (MRE).

Since the epoch of its phase transition to the present, LFDM redshifts identically to CDM, implying that
\begin{equation}
 \label{eq:y:1}
 \rho_{\rm LFDM}(z) =  \rho_{\rm LFDM}(z_T)\frac{(1+z)^3}{(1+z_T)^3},
\end{equation}
where $\rho_{\rm LFDM}(z)$ is the LFDM density evaluated at redshift $z$. Assuming that a fraction of excess radiation is converted into the entire CDM density at redshift $z_T$, this yields the following decrement in the effective number of neutrino degrees of freedom:
\begin{equation}
 \label{eq:y:2}
 \Delta N_{\rm eff} \rho_{\nu}(z_T) = \rho_{\rm LFDM}(0)(1+z_T)^3,
\end{equation}
where $\rho_{\nu}(z_T)$ is the energy density of one neutrinolike radiation species at the formation epoch. Thus, we have
 \begin{equation}
 \label{eq:y:3}
  \Delta N_{\rm eff}= \frac{\rho_{\rm CDM}(0)}{\rho_{\nu}(0)} \approx 0.2 \left( \frac{\Omega_{\rm CDM} h^2}{0.1199} \right) \left ({\frac{10^5}{1+z_T}} \right ).
\end{equation}
Note that $\Delta N_{\rm eff}$ is inversely proportional to the redshift
of the LFDM phase transition. Because the effective number of neutrino degrees of freedom changes dynamically in this model, observational constraints on $N_{\rm eff}$ must be interpreted with caution.

For most LFDM phase transition epochs between BBN and the CMB, the resulting value of $\Delta N_{\rm eff}$ is smaller than the precision of current observational constraints on this quantity; for instance, Equation \eqref{eq:y:3} implies that $z_T=10^5$ corresponds to~$\Delta N_{\rm eff}$ = 0.2, assuming the best-fit \emph{Planck} value of~$\Omega_{\rm CDM} h^2=0.1199$~\cite{2016A&A...596A.107P}. Recent constraints on $N_{\rm eff}$ from \emph{Planck} and WMAP prefer the existence of a fractional dark radiation component, with $\Delta N_{\rm eff} = 0.15$ at $95\%$ confidence \cite{Aghanim:2018eyx}. This bound is relaxed in the presence of nontrivial dark radiation self-interactions, which modify standard cosmological behavior during the radiation-dominated epoch \cite{Kreisch:2019yzn}. 
Thus, LFDM is in complete agreement with $\Delta N_{\rm eff}$ constraints if the phase transition occurs before $z \sim 10^5$, in which case $\Delta N_{\mathrm{eff}} \ll 0.2$ is sufficient to account for the entire DM relic density. Such a small fractional change in $\Delta N_{\mathrm{eff}}$ from an $\sim \rm{eV}$ neutrinolike particle also affects CMB density perturbations; in particular, modes with $\ell > 200$ that enter the horizon between BBN and the CMB respond to the presence of this tiny dark radiation excess. Constraints from this effect are compatible with the typical values of $\Delta N_{\mathrm{eff}}$ required for LFDM to constitute the entirety of DM \cite{Schoneberg:2019wmt}.

Importantly, unlike WIMPs (which couple to the SM through the weak interaction) or QCD axions (which primarily couple to the SM through electromagnetic interactions), LFDM need not have any interactions with the visible sector. Direct detection signatures are therefore not guaranteed for LFDM, although they are possible for specific constructions of the model. On the other hand, the suppression of the linear matter power spectrum, which manifests as a suppression of the power inferred from various tracers throughout cosmic history (e.g., \cite{Sarkar14107129,Corasaniti161105892}), is \emph{inevitable} in LFDM. In addition, dark acoustic oscillations (DAOs) imprinted prior to the phase transition can leave distinct signatures; for example, the $21$-cm brightness power spectrum may be enhanced in LFDM models relative to CDM \cite{Das:2017nub}.

\subsection{A.\ Self-interacting LFDM}
\label{model_si}

SI LFDM is a natural model in which the phase transition from a dark radiation component to a CDM state can easily be achieved. Recently, it has been shown that $\sim \mathrm{eV}$ sterile neutrinolike dark fermions, which have strong self-interactions mediated by a sub-$\mathrm{eV}$ scalar field, can be trapped into DM ``nuggets'' in the radiation-dominated era, a few $e$-foldings before the CMB~\cite{Das200511889}. The phase transition occurs when the attractive scalar fifth force overcomes free-streaming, which traps all of the $\sim \mathrm{eV}$ fermions within a Compton volume into degenerate DM nuggets. Collectively, these nuggets behave exactly like CDM and are produced with negligible thermal velocities due to their $\sim \mathrm{TeV}$ mass, unlike other LFDM models with non-negligible peculiar velocities that evolve ballistically after the phase transition \citep{Das181100028}. The stability of the nuggets is achieved by fermion degeneracy pressure, which balances the scalar fifth force, and the duration of the phase transition is negligible compared to the Hubble time for any transition redshift prior to the epoch of the CMB. Because of the heavy, composite nature of the nuggets resulting from their nonlinear formation process, the initial distribution function of the thermal dark fermions is not conserved. Thus, the nuggets avoid the Tremaine-Gunn phase-space bound derived from the internal dynamics of dwarf galaxies that applies to other light fermionic dark matter and WDM candidates \citep{Tremaine1979,Boyarsky08083902,Alvey201003572}. This model therefore provides a concrete construction of a phase transition in which a fluid that initially behaves like dark radiation changes its equation of state almost instantaneously at a transition redshift $z_{T,\mathrm{SI}}$.

Bosonic SI LFDM appears in theories of neutrino dark energy, in which neutrinos interact with multiple scalar fields and behave like a single thermalized fluid \cite{Das:2006ht}. In these theories, the scalars generally have hybrid potentials reminiscent of hybrid inflationary potentials. As the neutrino temperature dilutes near the epoch of MRE, one of the scalar fields that was stuck in a metastable minimum becomes tachyonic and begins to oscillate around a new minimum. The coherently oscillating field then behaves exactly like CDM, similar to the transition axion dark matter undergoes when the Hubble rate drops below its oscillation frequency. 

From a theoretical perspective, the epoch of the LFDM phase transition in neutrino dark energy theories is expected to be very late, and is therefore subject to constraints arising from linear perturbation theory. In particular, the relevant range of LFDM formation epochs can be estimated by assuming that the coupling of the particle model is of $\mathcal{O}(1)$, which yields~$1\ \mathrm{eV} \lesssim T(z_{T,\mathrm{SI}}) \lesssim 10^3\ \mathrm{eV}$ for the temperature of the Universe at the phase transition \cite{Das:2006ht}. The wave numbers corresponding to horizon entry for this range of transition epochs are $2\times 10^{-2} h\ \mathrm{Mpc}^{-1} \lesssim k_{T,\mathrm{SI}} \lesssim 20 h\ \mathrm{Mpc}^{-1}$. We reiterate that this is an order-of-magnitude estimate that only assumes natural values of the coupling constants.

\subsection{B.\ Free-streaming LFDM}

In the FS LFDM model, a noninteracting dark radiation component that free-streams until the DM phase transition starts to oscillate coherently and behave like CDM at redshift $z_{T,\mathrm{FS}}$. It is shown in \cite{Das:2006ht} that a thermal field theory correction can in principle make this phase transition possible. In particular, consider a scalar field $\phi$ with mass $m$ and a zero-temperature potential
\begin{equation}
 \label{thermal}
V(\phi)= V_0 - \frac{m^2 \phi^2}{2} -\epsilon \phi^3 + \frac{\lambda \phi^4}{4}, 
\end{equation}
where $V_0$ is the zero-point energy and $\epsilon$, $\lambda$ are coupling constants. This potential can pick up a correction due to the presence of other fermionic fields at finite temperature, resulting in fluctuations
\begin{equation}
\label{thermal correction}
\delta V = D T^2 \phi^2 \newline 
\end{equation}
where $D$ depends on the spin, coupling, and number of degrees of freedom of the other fields.

Here we have assumed that $\phi$ is not in thermal equilibrium with other fields, which implies that $\phi$ is noninteracting in a cosmological sense. With such a potential, the field is trapped in a minimum at $\phi=0$ for $T \geq m/\sqrt{2 D}$~\cite{Das:2006ht}. After the Universe cools below this temperature, the field becomes tachyonic about the origin and settles into the true minimum, after which it coherently oscillates and behaves like CDM. This model is therefore a concrete example of FS LFDM.


\section{III.\ Linear Perturbations}
\label{matter_pert}

\subsection{A.\ Free-streaming LFDM}

Despite the variety of particle models described above, the initial conditions for LFDM matter perturbations after its phase transition are identical to that of a dark radiation component at the transition epoch. If the dark radiation component has no self-interactions, then matter perturbations can be treated exactly as in the case of neutrinos, and the evolution of FS LFDM density perturbations is obtained by solving a series of coupled differential equations \cite{Ma9506072}:
\begin{eqnarray}
\dot{\delta} &=& -\frac{4}{3} \theta - \frac{2}{3} \dot h, \nonumber \\
\dot{\theta} &=& k^2\left(\frac{\delta}{4} - \sigma \right), \nonumber \\
2 \dot{\sigma} &=& \frac{8}{15} \theta - \frac{3}{15} k F_3
+ \frac{4}{15} \dot h + \frac{8}{5} \dot{\eta},\ \mathrm{and} \nonumber \\
\dot{F}_\ell &=& \frac{k}{2\ell+1} \left(\ell F_{\ell-1} - (\ell+1) F_{\ell+1}
\right),
\label{eq:series}
\end{eqnarray}
where $\delta$ is the LFDM overdensity field, $\theta$ is its velocity divergence, $h$ and $\eta$ are metric perturbations in synchronous gauge, $\sigma$ is the shear stress, $F_{\ell}$ is the $\ell$th Legendre component of the momentum-averaged LFDM distribution function, $k$ is the cosmological wave number, and overdots denote derivatives with respect to conformal time \cite{Ma9506072}. The solution for $\delta$ is an exponentially damped oscillator at subhorizon scales; physically, this represents the free-streaming of highly relativistic neutrinos.

To compute the growth of linear matter perturbations for the FS LFDM model, we modify the Boltzmann solver CAMB to evolve matter fluctuations up to a redshift $z_{T,\mathrm{FS}}$ without CDM, and we extract the transfer function for neutrino perturbations at this redshift according to Eq.~\eqref{eq:series}. We then use these neutrino (dark radiation) perturbations as initial conditions for LFDM density fluctuations at the epoch of its formation, and we evolve LFDM perturbations identically to CDM thereafter to obtain the linear matter power spectrum at later times. Thus, oscillations at small scales in the linear matter power spectrum arise because LFDM obtained its initial density fluctuations from neutrinolike perturbations at $z_{T,\mathrm{FS}}$, which were damped and oscillatory at scales smaller than the size of the horizon at that time.

\subsection{B.\ Self-interacting LFDM}

Equation \eqref{eq:series} provides the initial conditions for a neutrinolike particle that transitions to CDM. For SI LFDM, the situation is simplified because a strongly self-interacting neutrinolike fluid can be treated in the tight-coupling approximation, in which the anisotropic stress and higher-order terms are neglected (analogous to the treatment of the photon-baryon fluid). The following equations then describe linear perturbations for the SI LFDM model: 
\begin{eqnarray}
\dot{\delta} & = & -\frac{4}{3} \theta - \frac{2}{3} \dot h, \nonumber \\
\dot{\theta} & = & k^2\left(\frac{\delta}{4} - \sigma \right).\label{eq:si_evolution}
\end{eqnarray}
We note that the above perturbation equations for a tightly coupled dark matter-radiation fluid are only valid until then epoch of the phase transition, and that---once LFDM forms---it behaves identically to cold, collisionless CDM. In our modified CAMB implementation, we therefore evolve matter perturbations until the redshift of the phase transition, $z_{T,\mathrm{SI}}$, according to Eq.~\eqref{eq:si_evolution}. We then use the solution as the initial condition for subsequent evolution, which is identical to CDM.


\section{IV.\ Transfer Functions}
\label{sec:transfer}

To compare linear matter power spectra in our LFDM models to CDM, we compute the \emph{transfer function}
\begin{equation}
    T^2(k)\equiv \frac{P_{\mathrm{LFDM}}(k)}{P_{\mathrm{CDM}}(k)},
\end{equation}
where $P_{\mathrm{LFDM}}(k)$ [$P_{\mathrm{CDM}}(k)$] is the LFDM (CDM) linear matter power spectrum evaluated at $z=0$. The \emph{half-mode scale} $k_{\mathrm{hm}}$ is defined as the wave number at which~$T^2(k) = 0.25$. 

\begin{figure*}[t]
\hspace{-4mm}
\includegraphics[scale=0.41]{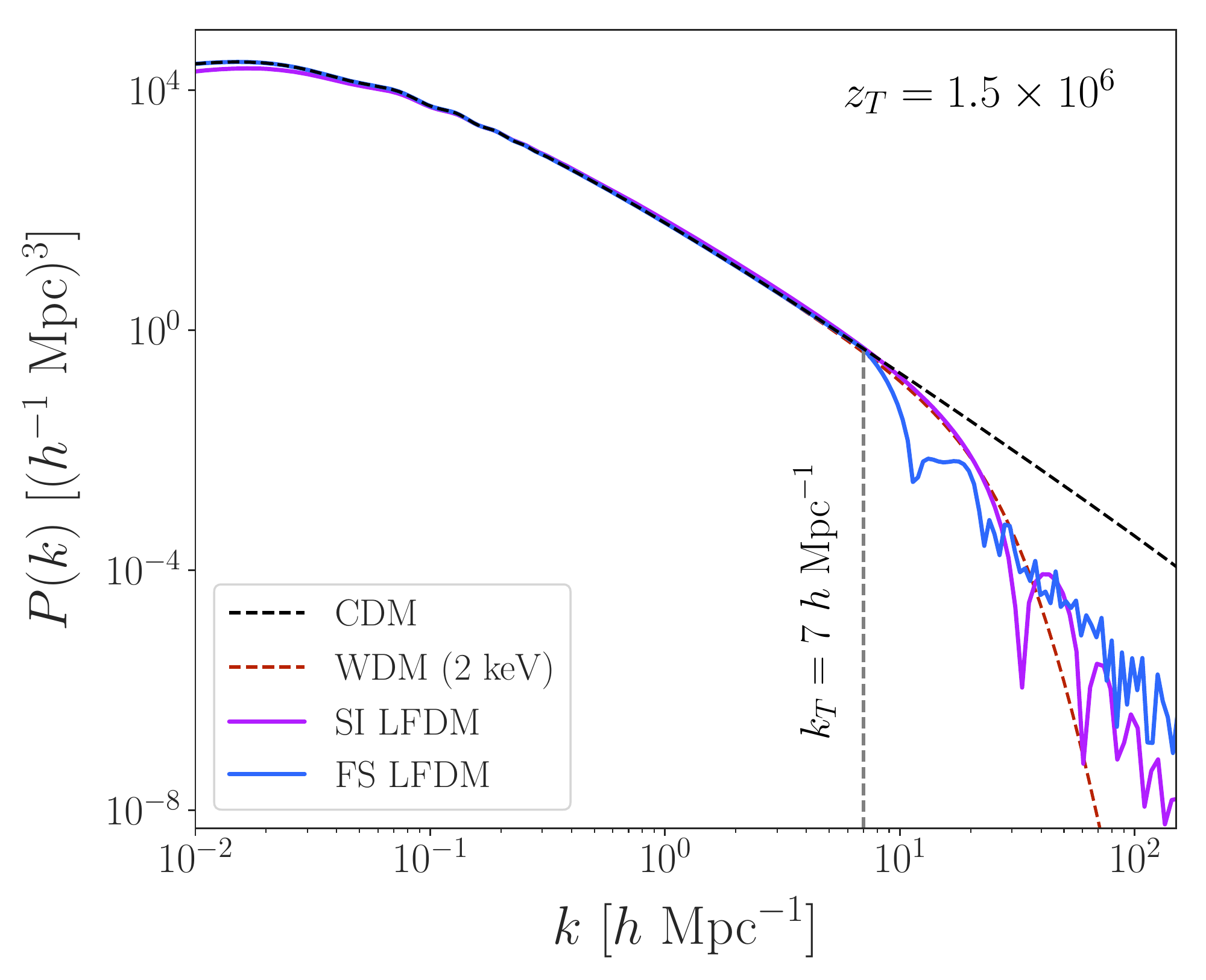}
\hspace{2mm}
\includegraphics[scale=0.41]{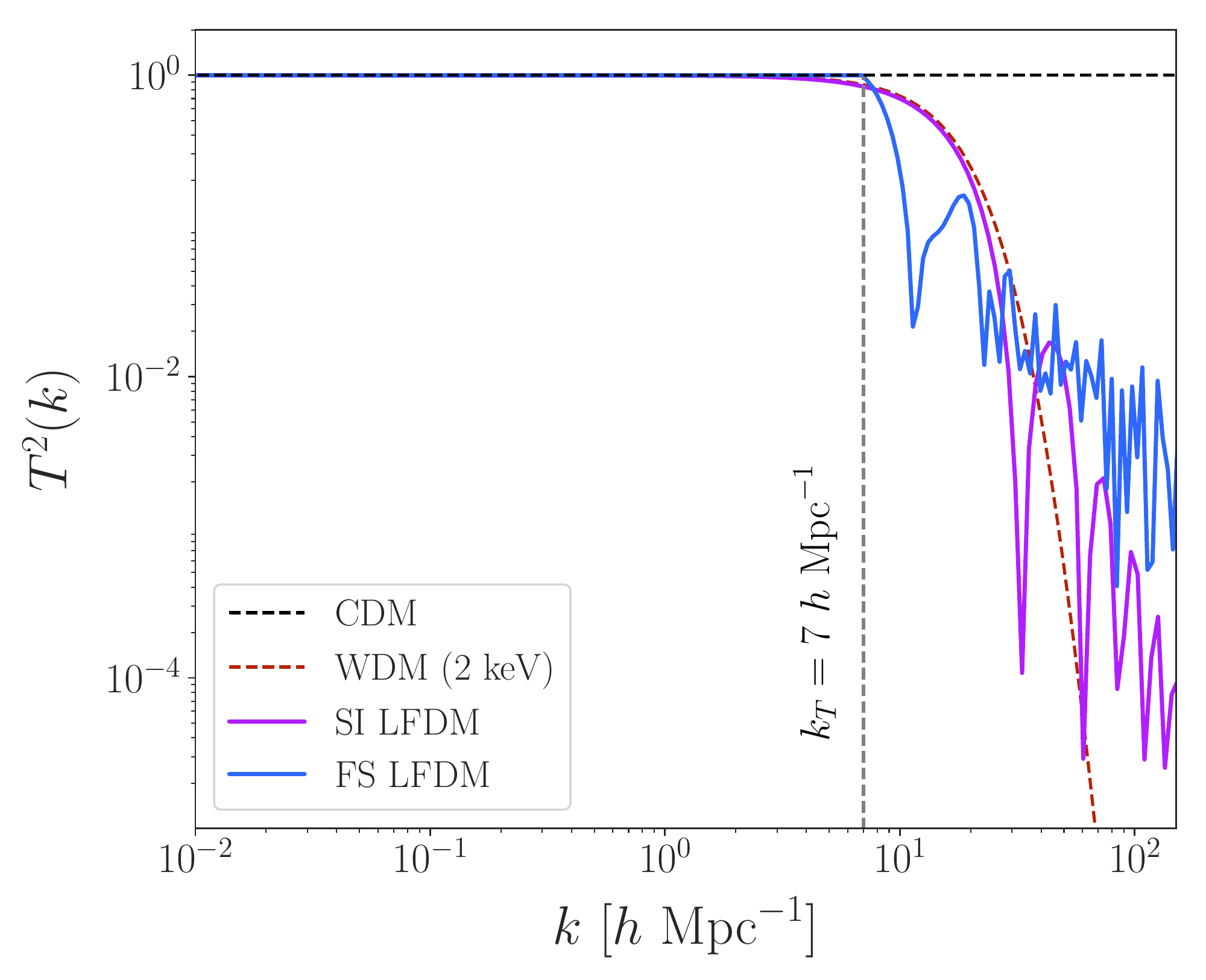}
\vspace{-1mm}
\caption{Linear matter power spectra (left) and transfer functions (right) for self-interacting (magenta) and free-streaming (cyan) late-forming dark matter models, compared to cold dark matter (dashed black) and thermal relic warm dark matter (dashed red). Both LFDM models are shown with a transition redshift of $z_T = 1.5\times 10^6$, corresponding to a comoving wave number of $k_T = 7h\ \mathrm{Mpc}^{-1}$. LFDM power spectra are suppressed relative to CDM at wave numbers greater than $k_T$, and they exhibit dark acoustic oscillations on even smaller scales, beginning at $\sim 6k_T$ ($\sim 2k_T$) for SI (FS) LFDM. The cutoff in the SI LFDM power spectrum is very similar to that in WDM, until the onset of DAOs.
\label{fig:pk_example}}
\end{figure*}

Linear matter power spectra and transfer functions for our SI and FS LFDM models with~$z_T = 1.5\times 10^6$ ($k_T = 7h\ \mathrm{Mpc}^{-1}$) are shown in Fig.~\ref{fig:pk_example}. We note that the transition redshift shown in Fig.~\ref{fig:pk_example} is marginally consistent with Lyman-$\alpha$ forest and galaxy clustering data \cite{Sarkar14107129}; however, as we demonstrate below, it is robustly ruled out for both LFDM models by our MW satellite population analysis. 

The right panel of Fig.~\ref{fig:pk_example} illustrates three main features of LFDM transfer functions that are common to both of our model variants:
\begin{enumerate}
    \item There is a cutoff in power relative to CDM at the comoving wave number $k_T$, which corresponds to the size of the horizon at the epoch of the LFDM phase transition. In particular, power is significantly suppressed on scales smaller than those corresponding to
    \begin{equation}
    k_T = \frac{aH_T}{c} \approx \frac{H_0\sqrt{\Omega_{\mathrm{rad}}}z_T}{c},\label{eq:transition}
    \end{equation}
    where $H_T$ is the Hubble rate at the LFDM transition, $H_0=100h\ \mathrm{km\ s}^{-1}\ \mathrm{Mpc}^{-1}$ is the present-day Hubble rate, and $\Omega_{\mathrm{rad}}\approx 10^{-4}$ is the energy density in radiation.\footnote{As discussed above, CMB constraints on $\Delta N_{\rm eff}$ set a limit of~$z_T\gtrsim 4\times 10^5$. Later transitions also result in severe suppression of the matter power spectrum on quasilinear scales according to Eq.~\eqref{eq:transition}.}
    \item There are damped DAOs at scales smaller than those corresponding to $k_T$, resulting from dark radiation perturbations prior to the LFDM phase transition.
    \item Cutoffs in the transfer functions for both model variants exhibit \emph{$k$-translation invariance}. Specifically, given two SI or FS LFDM models with transition redshifts $z_{T,1}$ and $z_{T,2}$ and transfer functions $T^2_1(k)$ and $T^2_2(k)$, we have
    \begin{equation}
        T_2^2(k) = T_1^2\left(\frac{z_{T,2}}{z_{T,1}}k\right)\label{eq:k_invariance}
    \end{equation}
    along the initial cutoff. This symmetry follows from the linear relation between $k_T$ and $z_{T}$ in Eq.~\eqref{eq:transition} and from the scale invariance of Hubble expansion in the radiation-dominated epoch. We emphasize that Eq.~\eqref{eq:k_invariance} only holds along the initial power spectrum cutoff; this is sufficient for our purposes because DAOs occur at extremely small scales for the typical transition redshift values we consider. Equation \eqref{eq:k_invariance} is useful because it allows us to analytically compute LFDM transfer functions as a continuous function of $z_{T}$ using the power spectra that were computed with CAMB for discrete transition redshifts.
    \end{enumerate}

\subsection{A.\ Self-interacting LFDM}

The SI LFDM transfer function exhibits a smooth cutoff that is remarkably similar to that in thermal relic WDM until the onset of DAOs. The tight correspondence between the cutoff in these transfer functions is reminiscent of the mapping between thermal relic WDM and velocity-independent DM-proton scattering found in~\cite{Nadler190410000}, and (to a lesser extent) a similar mapping identified for models with DM-radiation interactions \cite{Boehm14047012,Escudero180308427}. Despite different dark matter microphysics, the transfer function for our SI LFDM model is also similar to that for self-interacting dark matter models in which massive dark photon mediators decay to dark fermions \citep{Huo170909717}. More generally, \cite{Boehm0112522,Boehm0410591} have shown that interacting DM models often impact the linear matter power spectrum such that they are effectively ``warm.'' The existence of the mapping between SI LFDM and thermal relic WDM is therefore not surprising given its strong self-interactions prior to the phase transition.

To make this correspondence quantitative, we construct a relation between the SI LFDM and thermal relic WDM models following a half-mode scale matching procedure similar to \cite{Escudero180308427,Nadler190410000}. In particular, we derive the following relation from our CAMB output:
\begin{equation}
    k_{\mathrm{hm},\mathrm{SI}} \approx 2.8k_{T,\mathrm{SI}} 
    \approx 1.3\left(\frac{z_{T,\mathrm{SI}}}{10^{5}}\right)h\ \mathrm{Mpc}^{-1}.
\end{equation}
Meanwhile, the half-mode scale in WDM is given by \cite{Viel0501562}
\begin{align}
    &k_{\mathrm{hm},\mathrm{WDM}} = \frac{2\pi}{\lambda_{\mathrm{hm},\mathrm{WDM}}}&\nonumber \\ 
    &= 9.2\left(\frac{m_{\mathrm{WDM}}}{1\ \mathrm{keV}}\right)^{1.11}\left(\frac{\Omega_{m}}{0.25}\right)^{-0.11}\left(\frac{h}{0.7}\right)^{-1.22}h\ \mathrm{Mpc}^{-1},&
\end{align}
where $m_{\mathrm{WDM}}$ is the thermal relic WDM mass. Solving for the transition redshift that causes the half-mode scales of the WDM and SI LFDM transfer functions to match yields the relation
\begin{equation}
    z_{T,\mathrm{SI}} \approx 7\times 10^5 \left(\frac{m_{\mathrm{WDM}}}{1\ \mathrm{keV}}\right)^{1.11}\left(\frac{\Omega_{m}}{0.25}\right)^{-0.11}\left(\frac{h}{0.7}\right)^{-1.22}.\label{eq:mapping}
\end{equation}
We find that LFDM and WDM transfer functions matched in this way agree to better than $\sim 5\%$ along the initial cutoff over the entire SI LFDM parameter space of interest.

Examples of SI LFDM transfer functions along with matched WDM transfer functions are shown in the left panel of Fig.~\ref{fig:fs_example}. On this plot, we indicate the comoving wave number corresponding to the \emph{minimum halo mass}, i.e., the lowest-mass halo inferred to host MW satellite galaxies. In particular, from an analysis of the MW satellite population using DES and PS1 data over nearly three-fourths of the sky, \cite{Nadler191203303} found that the lowest peak virial halo mass corresponding to observed MW satellite galaxies is less than~$\mathcal{M}_{\mathrm{min}}=3.2\times 10^8\ \mathrm{M}_{\mathrm{\odot}}$ at $95\%$ confidence, corresponding to a comoving wave number of $k_{\mathrm{crit}}\approx 36h\ \mathrm{Mpc}^{-1}$. We also indicate the WDM transfer function ruled out by these observations of the MW satellite population at $95\%$ confidence, corresponding to a $6.5\ \mathrm{keV}$ thermal relic \cite{Nadler200800022}.

\subsection{B.\ Free-streaming LFDM}

The power spectrum cutoff in FS LFDM is significantly sharper than in SI LFDM, as expected due to its free-streaming behavior prior to the phase transition. Thus, it is difficult to directly map FS LFDM to WDM, which forces us to take a more conservative approach in order to derive constraints.

\begin{figure*}[t]
\hspace{-6mm}
\includegraphics[scale=0.415]{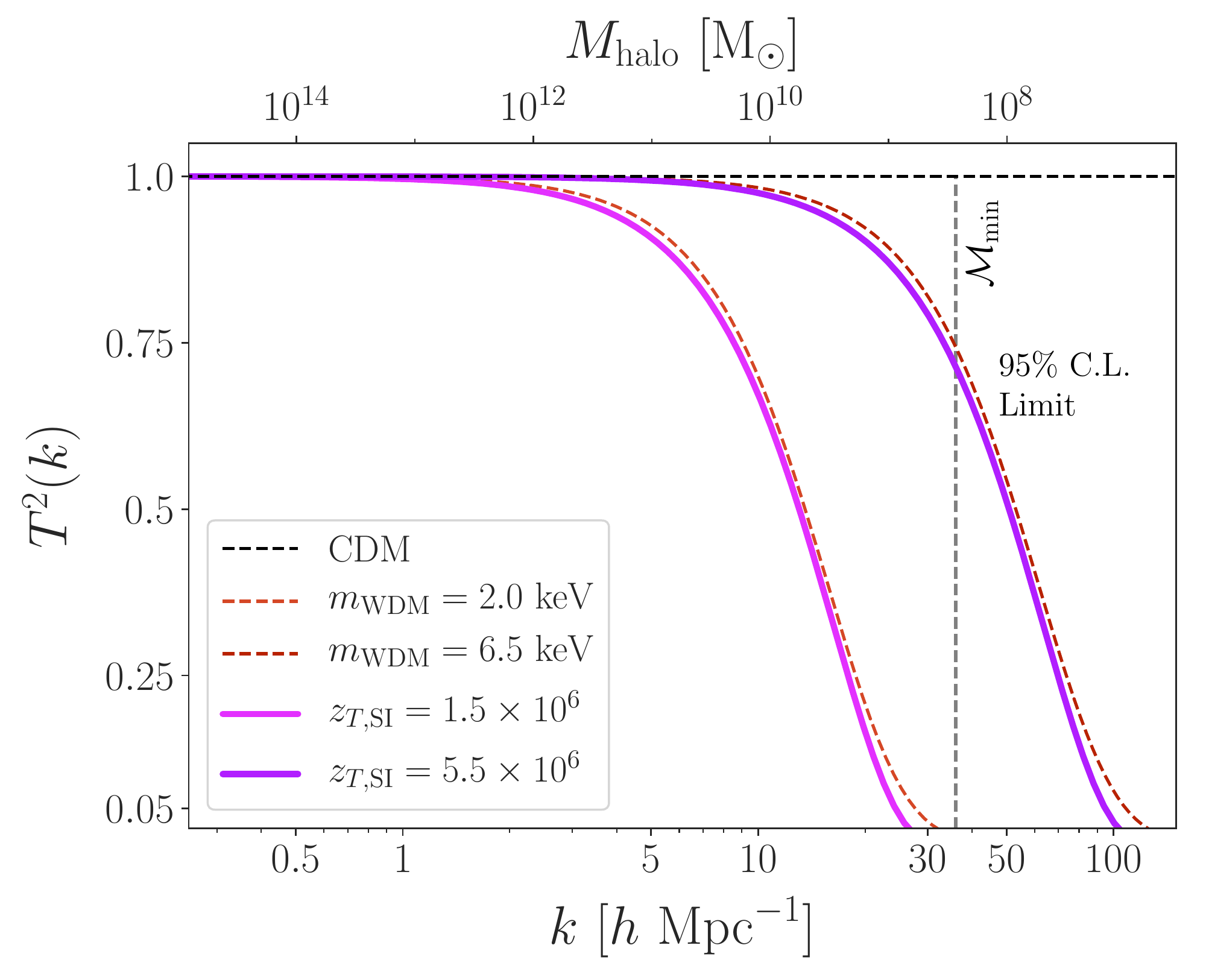}
\hspace{1.5mm}
\includegraphics[scale=0.415]{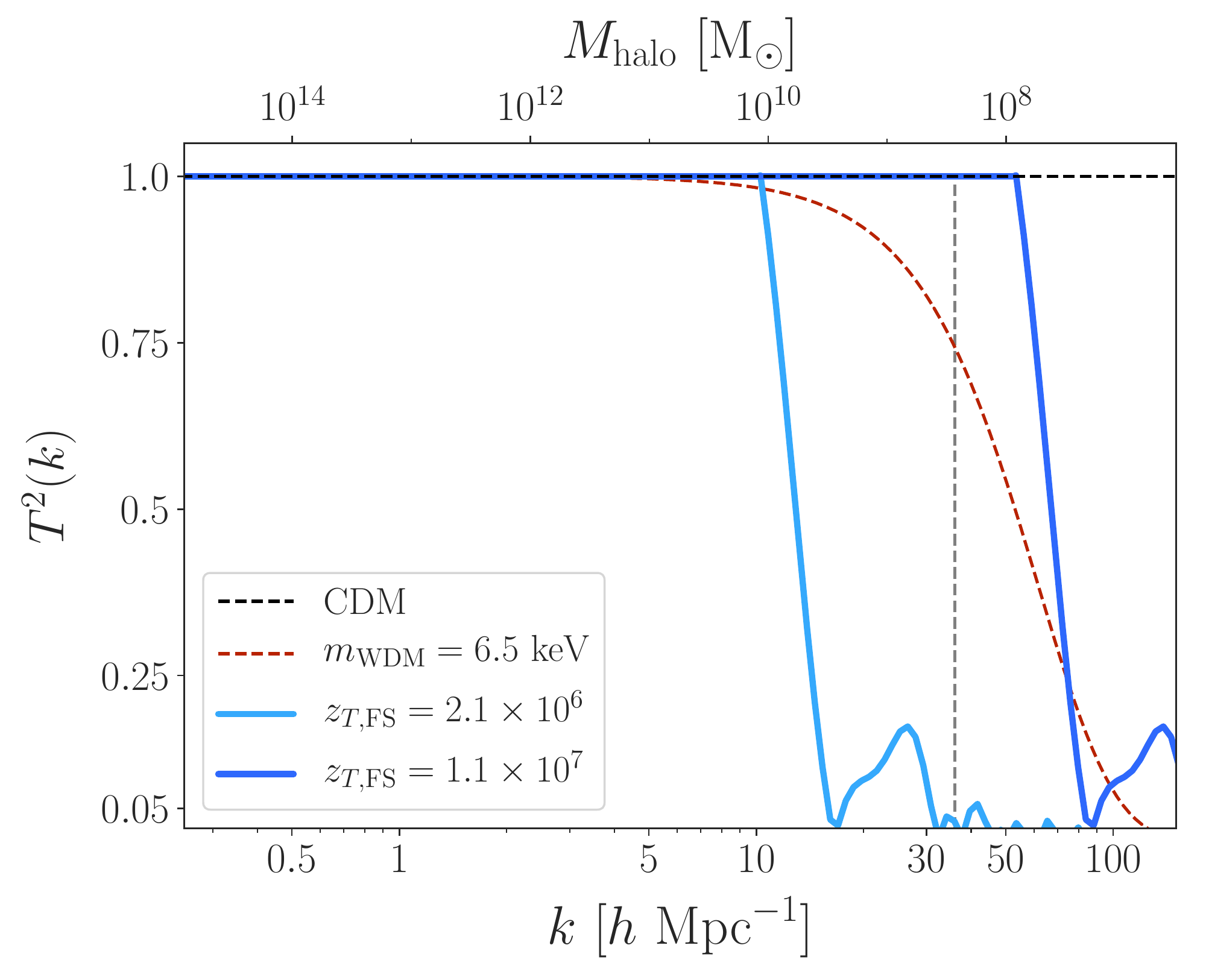}
\vspace{-1mm}
\caption{Transfer functions for self-interacting (left) and free-streaming (right) late-forming dark matter models, compared to cold dark matter (dashed black) and thermal relic warm dark matter (dashed red). SI LFDM models are shown for a range of transition redshifts, with the highest transition redshift corresponding to the SI LFDM model that is ruled out by the abundance of Milky Way satellites at $95\%$ confidence: $z_{T,\mathrm{SI}}>5.5\times 10^6$. The light-blue FS LFDM model corresponds to the transition redshift that is conservatively ruled out by our analysis:~$z_{T,\mathrm{FS}}>2.1\times 10^6$. Vertical dashed lines show the comoving scale that approximately corresponds to the mass of the smallest halo inferred to host observed MW satellite galaxies, $3.2\times 10^8\ \mathrm{M}_{\mathrm{\odot}}$ \cite{Nadler191203303}. In the left panel, WDM transfer functions are slightly shifted horizontally for visual clarity.
\label{fig:fs_example}}
\end{figure*}

Nonetheless, we can still construct a relation between the half-mode scale and the transition redshift for FS LFDM based on our CAMB output. This yields
\begin{equation}
    k_{\mathrm{hm},\mathrm{FS}} \approx 1.4k_{T,\mathrm{FS}} \approx 0.65\left(\frac{z_{T,\mathrm{FS}}}{10^{5}}\right)h\ \mathrm{Mpc}^{-1}.
\end{equation}
For a fixed transition redshift, $k_{\mathrm{hm,FS}}<k_{\mathrm{hm,SI}}$, which makes sense given the sharper power spectrum cutoff in FS LFDM relative to SI LFDM. FS LFDM transfer functions are shown in the right panel of Fig.~\ref{fig:fs_example}.

\begin{figure}[t]
\hspace{-3.75mm}
\includegraphics[scale=0.39]{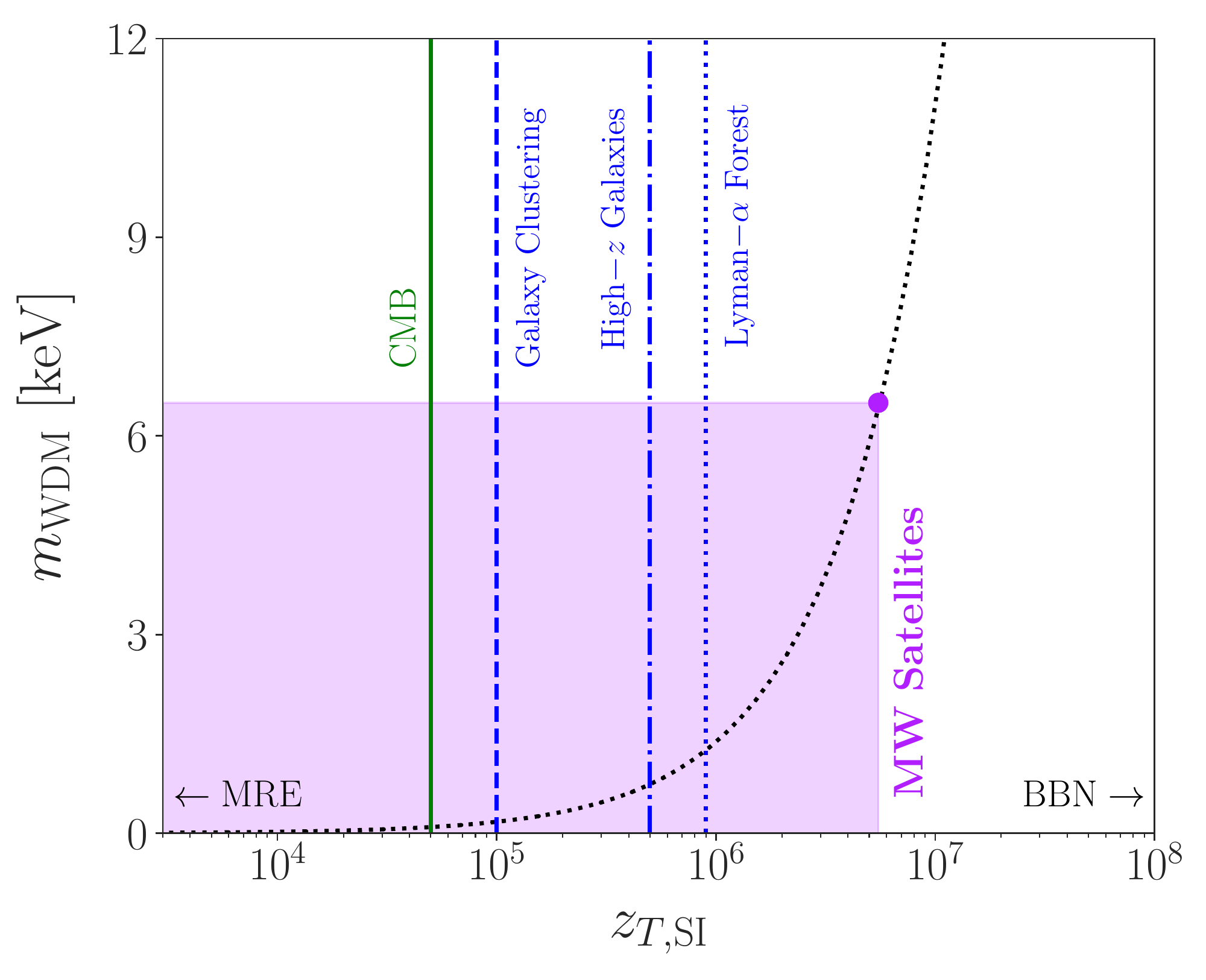}
\vspace{-2mm}
\caption{Constraints on the transition redshift for self-interacting late-forming dark matter, versus the corresponding thermal relic warm dark matter mass based on the half-mode mass relation in Eq.~\eqref{eq:mapping}. Our Milky Way satellite constraint on $z_{T,\mathrm{SI}}$ and the lower limit on the thermal relic WDM mass of $6.5\ \mathrm{keV}$ from which we derive this limit \cite{Nadler200800022} are shown by the shaded purple region. Limits on the SI LFDM transition redshift from the cosmic microwave background (green), Sloan Digital Sky Survey galaxy clustering (dashed blue \cite{Sarkar14107129}), the high-redshift galaxy luminosity function (dot-dashed blue \cite{Corasaniti161105892}) and the Lyman-$\alpha$ forest (dotted blue \cite{Sarkar14107129}) are shown as vertical lines. Vertical lines indicate constraints derived specifically for LFDM, and do not indicate other recent WDM constraints from small-scale structure probes. LFDM must transition to CDM between matter-radiation equality ($z\approx 3\times 10^3$) and big bang nucleosynthesis ($z\approx 10^{10}$), which are schematically indicated by arrows.
\label{fig:zt_wdm}}
\end{figure}


\section{V.\ Constraints from Milky Way Satellites}
\label{results}

We use the relations derived above to translate thermal relic WDM limits from the MW satellite population into LFDM constraints. Given that halos with masses lower than $3.2\times 10^{8}\ \mathrm{M}_{\mathrm{\odot}}$ are required to host currently observed MW satellite galaxies \cite{Nadler191203303}, there must be enough power to form bound DM halos on the corresponding comoving scales---i.e., down to a critical wave number of
\begin{equation}
    k_{\mathrm{crit}} = \frac{2\pi}{\lambda_{\mathrm{min}}} = \pi \left(\frac{4\pi\rho_m}{3\mathcal{M}_{\mathrm{min}}}\right)^{1/3}\approx 36h\ \mathrm{Mpc}^{-1},
\end{equation}
where $\rho_m$ is the LFDM density today, $\mathcal{M}_{\mathrm{min}}$ is the minimum halo mass, and $\lambda_{\mathrm{min}}$ is the corresponding length scale in linear theory. Halos at this mass scale need not merely exist, but must be formed in enough \emph{abundance} to match the observed MW satellite population population. Thus, we will obtain a lower limit on the transition redshift in both LFDM models based on the lower limit on the thermal relic WDM mass.

\subsection{A.\ Self-interacting LFDM}

The LFDM-WDM mapping constructed above allows us to translate thermal relic WDM limits derived from the MW satellite population into LFDM constraints. High-resolution cosmological simulations have been performed in order to predict the WDM subhalo mass function in MW-mass halos \cite{Schneider11120330,Lovell13081399,Angulo13042406,Bose160407409}, and these have been used in conjunction with the observed MW satellite population to place stringent constraints on thermal relic WDM. \cite{Nadler200800022} report $m_{\mathrm{WDM}}>6.5\ \mathrm{keV}$ at $95\%$ confidence, which we directly translate into a constraint on SI LFDM via Eq.~\eqref{eq:mapping}, yielding $z_{T,\mathrm{SI}}>5.5\times 10^{6}$, also at~$95\%$ confidence. This limit implies that the dark radiation which transitions to LFDM causes $\Delta N_{\rm eff}\lesssim 4\times 10^{-3}$, assuming that LFDM constitutes the entire DM relic density [Eq.~\eqref{eq:y:3}]. Exploring the generality of this indirect constraint on $\Delta N_{\rm eff}$ from small-scale structure measurements is a compelling avenue for future work.

Figure \ref{fig:zt_wdm} compares this limit to constraints on $z_{T,\mathrm{SI}}$ derived from the CMB (resulting from $N_{\mathrm{eff}}$ constraints), low-redshift galaxy clustering from the Sloan Digital Sky Survey \cite{Sarkar14107129}, the high-redshift galaxy luminosity function~\cite{Corasaniti161105892}, and the Lyman-$\alpha$ forest \cite{Sarkar14107129}. Our limit improves upon the Lyman-$\alpha$ forest result by a factor of $\sim 6$, which can be understood in terms of the comoving scales probed by the MW satellite population. Specifically, the lowest-mass halo inferred to host an observed satellite is $\sim 3\times 10^8\ \mathrm{M}_{\mathrm{\odot}}$ \cite{Nadler191203303}, which roughly corresponds to a wave number of~$k\sim 40h\ \mathrm{Mpc}^{-1}$, while the Lyman-$\alpha$ forest data used in~\cite{Sarkar14107129} reaches $k\sim 5h\ \mathrm{Mpc}^{-1}$. We expect $z_{T,\mathrm{SI}}$ to scale linearly with the wave number corresponding to the smallest scale probed in an observational analysis, and the improvement we observe relative to this Lyman-$\alpha$ constraint is consistent with this expectation.\footnote{More recent Lyman-$\alpha$ forest analyses (e.g., \cite{Rogers200712705,Rogers200713751}) probe smaller scales and a wider range of redshifts, and will therefore improve upon the LFDM constraints in \cite{Sarkar14107129}.} Other small-scale probes that achieve comparable sensitivity to thermal relic WDM, including strong gravitational lensing~\cite{Gilman190806983,Hsueh190504182} and stellar streams \cite{Banik191102662}, will yield similar LFDM constraints.

Our SI LFDM limit relies on an analytic mapping to thermal relic WDM and is therefore not directly validated using LFDM simulations. We note that \cite{Corasaniti161105892} ran simulations of these models with similar half-mode scales and found that the high-redshift ($z>4$) LFDM halo mass function is comparable to that in WDM. Those findings are further consistent with the suite of LFDM simulations from \cite{Agarwal14121103}, which show that oscillatory features in the linear matter power spectrum are erased in the $z=0$ halo mass function. Meanwhile, \cite{Bohr200601842}---working in the Effective Theory of Structure Formation (ETHOS) framework \cite{Cyr-Racine151205344}---found the peak heights of interest for our SI LFDM constraints lead to negligible differences in the high-redshift halo mass function relative to thermal relic WDM. Finally, \cite{Huo170909717} showed that the halo mass function for self-interacting dark matter models with similar transfer functions to our SI LFDM model are nearly indistinguishable from matched WDM models, and used this correspondence along with a conservative treatment of the subhalo population inferred from MW satellites to place constraints similar in spirit to ours. All of these results lend confidence to the robustness of our result when framed as a conservative limit.

\subsection{B.\ Free-streaming LFDM}

The right panel of Fig.~\ref{fig:fs_example} demonstrates the reason that it would be dangerous to set a constraint on FS LFDM based on matching its half-mode scale to WDM. In particular, because the FS LFDM power spectrum cutoff is much steeper than in thermal relic WDM, the half mode-matched model is significantly \emph{less} suppressed than the corresponding WDM model along the initial power spectrum cutoff. Thus, we bracket the range of allowed FS LFDM transition redshifts as follows:
\begin{enumerate}
    \item We place a \emph{fiducial} lower limit on $z_{T,\mathrm{FS}}$ by finding the FS LFDM transfer function that yields strictly greater power suppression than the ruled-out thermal relic WDM model for all wave numbers~$k>10h\ \mathrm{Mpc}^{-1}$, roughly corresponding to halo masses below $10^{10}\ \mathrm{M}_{\mathrm{\odot}}$.\footnote{This procedure is similar to that used to constrain resonantly produced sterile neutrinos in \cite{Schneider160107553,Nadler200800022} and developed by \cite{Maamari} to constrain velocity-dependent DM-proton interactions.} Below this wave number, small differences between the FS LFDM and WDM transfer functions are negligible for the FS LFDM models of interest. This yields a conservative limit of $z_{T,\mathrm{FS}}>2.1\times 10^6$ and is shown by the light-blue transfer function in Fig.~\ref{fig:fs_example}.
    \item We forecast an \emph{optimistic} limit on $z_{T,\mathrm{FS}}$ by matching it to the half-mode scale of the thermal relic WDM model that is ruled out at $95\%$ confidence by the MW satellite population. This yields~$z_{T,\mathrm{FS}}>1.1\times 10^7$ and is shown by the dark-blue transfer function in Fig.~\ref{fig:fs_example}. This constraint is optimistic because the abundance of subhalos that host MW satellites are sensitive to a convolution of power on (nonlinear) scales, rather than a single mode at which the power spectrum is suppressed by a characteristic amount (e.g., $k_{\mathrm{hm}}$); thus, transfer functions with different cutoff shapes cannot be matched in detail.
\end{enumerate}

Because the FS LFDM model has not previously been considered in the context of small-scale structure measurements, we do not have a direct point of comparison for our constraints on its transition redshift. However, our \emph{fiducial} FS LFDM is extremely conservative. It is therefore clear that $z_{T,\mathrm{FS}}$ must be of the same order-of-magnitude $z_{T,\mathrm{SI}}$, which is physically reasonable.

Like our SI LFDM constraint, our forecasted \emph{optimistic} limit on $z_{T,\mathrm{FS}}$ is analytic and therefore must be confirmed with measurements of the subhalo mass function in dedicated LFDM simulations of MW-like systems. This situation is reminiscent of that for fuzzy dark matter (FDM), which also features steeper power suppression (for a fixed half-mode scale) than thermal relic WDM. Half-mode matching predicts a stringent limit on the FDM mass (e.g., \cite{Nadler190410000}); however, constraints based directly on the FDM subhalo mass function are less strict \cite{Schutz2001055503,Nadler200800022}. We are therefore confident that the correct limit on $z_{T,\mathrm{FS}}$ lies between our \emph{fiducial} and \emph{optimistic} constraints.


\section{VI.\ Conclusion}
\label{conclusion}

In this study, we set novel constraints on the dark matter formation epoch using state-of-the-art limits on the suppression of the small-scale matter power spectrum from the Milky Way satellite population. Specifically, we focused on the theoretically motivated paradigm of late-forming dark matter, which transitions to collisionless, cold dark matter from a dark radiation state. We showed that the epoch of the LFDM transition determines the cutoff scale in the linear matter power spectrum, which is processed into a suppression of power throughout cosmic history. By exploiting the correspondence between the power spectrum cutoff in a LFDM model with strong self-interactions prior to the phase transition versus that in thermal relic warm dark matter, we used the latest WDM constraint from the MW satellite population to place a stringent lower limit on the LFDM transition redshift. This constraint improves upon previous results by nearly an order of magnitude. We also estimated lower limits on the transition redshift for free-streaming LFDM.

Crucially, several independent tracers of small-scale structure corroborate the dark matter constraints set by recent MW satellite studies; thus, our constraints are not highly dependent on the particular probe used to set the WDM limit we exploited in this paper. In particular, analyses of the Lyman-$\alpha$ forest flux power spectrum \cite{Viel:2013,Irsic:2017}, strongly lensed quasar flux ratio anomalies and magnifications~\cite{Hsueh190504182,Gilman190806983}, and perturbations in Galactic stellar streams \cite{Banik191102662} have achieved similar sensitivity to thermal relic WDM relative to the MW satellite population, even though the observational and theoretical systematics of these probes differ. Thus, these other small-scale structure probes can also be used to constrain the dark matter transition redshift. This is particularly important because the dark acoustic oscillations imprinted prior to the LFDM phase transition can potentially have distinct consequences for different tracers of the matter power spectrum at various epochs (e.g., \cite{Das:2017nub}).

Extending the sensitivity of dark matter formation epoch measurements to even earlier times requires probing the linear matter power spectrum on extremely small scales. For example, ruling out the possibility that LFDM forms after BBN requires sensitivity to linear modes with $k \sim 10^5h\ \mathrm{Mpc}^{-1}$, or halos with masses of~$\sim 10^{-2}\ \mathrm{M}_{\mathrm{\odot}}$. These tiny, baryon-free halos are only detectable through their gravitational effects, which next-generation pulsar timing arrays \cite{Ramani200503030} and gravitational wave lensing measurements \cite{Oguri200701936} can potentially discover.


\section{Acknowledgments}

We thank Arka Banerjee, Keith Bechtol, Anirban Das, Alex Drlica-Wagner, and Risa Wechsler for comments on the manuscript, and Kimberly Boddy and Marc Kamionkowski for initial discussions. S.\ D.\ acknowledges the IUSSTF-JC-009-2016 grant from the Indo-U.S.\ Science \& Technology Forum which supported the travel where the project was initiated. 
This research received support from the National Science Foundation (NSF) under Grant No.\ NSF DGE-1656518 through the NSF Graduate Research Fellowship received by E.\ O.\ N. 


\bibliography{bibliography1}{}

\end{document}